# Lithium generated by cosmic rays: an estimator of the time that Mars had a thicker atmosphere and liquid water

Hector Javier Durand-Manterola

Space Science Department, Institute of Geophysics, National Autonomous University of Mexico

hdurand_manterola@yahoo.com

*Abstract*
Lithium is overabundant in cosmic rays because protons impact on carbon and oxygen nuclei and fission them. Among the products of this fission is lithium. Given this preference for carbon and oxygen atoms, in this work I propose that in an atmosphere of almost pure $CO_2$, such as Mars and Venus atmospheres, lithium nuclei are produced by interaction with cosmic rays. I calculated the production rate of lithium and came to the conclusion that, for pressures of two bars or greater, are produced between 21 and 81 lithium nuclei for each primary cosmic rays proton. For lower pressures, the production is less and almost nil with the current pressure of Mars or Earth (pressure of $CO_2$). Assuming a rate of cosmic ray arrival at Mars equal to that of Earth, and a pressure greater than two bars throughout the history of Mars, the amount of lithium that would occur would be between 162 and 642 million metric tons (in the Earth lithium estimated reserves are 30 million metric tons). These values are an upper limit; the actual amount of lithium on Mars will depend on the time in which the planet had a dense atmosphere (> 2 bars). That is, the amount of lithium produced by cosmic rays, serves to estimate the time that Mars had a thick atmosphere and therefore the capacity for have liquid water on surface.

*1 Introduction*
Lithium is the third element of the periodic table. Their abundance in the Sun is very low, 4.28 Li atoms per $10^6$ atoms of Si (Clayton, 2003). In general, solar abundances are taken as universal abundances considering the Sun a typical star. Lithium has two isotopes, lithium 6, with three neutrons in its nucleus and lithium 7 with 4 neutrons. Since it is a very reactive element is never isolated. The lithium forms, carbonates, carbides, nitrates and borates. Some of these compounds in aqueous solution lower the freezing point of water.
On Earth about 60% of total reserves of lithium are found in brines (Lagos-Miranda, 2009) because the lithium compounds are very soluble in water. In seawater is at concentrations of 0.17 ppm (Lenntech, 2012).

# Lithium generated by cosmic rays

In cosmic rays there is an overabundance of lithium. This is because particles of cosmic rays strike the interstellar gas atomic nuclei and fragmented forming lighter nuclei. The main atoms, which when struck form lithium, are carbon and oxygen (Clayton, 2003). If these atoms are the main producers of lithium in space then an atmosphere composed mainly of $CO_2$ has a high probability to generate this element, when struck by cosmic rays.

It is the thesis of this paper that the presence of an atmosphere of $CO_2$ on a planet causes the formation of lithium by interaction with cosmic rays on the same atmosphere of the planet. For this reason, on planets with this feature, like Mars and Venus, must have more lithium than on planets with low abundance of $CO_2$ in the atmosphere, as in the case of the Earth.

*2. Lithium generated by cosmic rays*
*2.1 Amount of cosmic rays that have impacted Mars*
The energy spectrum of cosmic rays outside the Earth's atmosphere is (Dorman, 2006, p 20)

$$n(E) = 8.6 \times 10^3 E^{-2.55} \qquad (1)$$

Where $n(E)$ is the number of particles per $m^{-2}$ $ster^{-1}$ $s^{-1}$ $Gev^{-1}$, and E is the energy.

If we want to know how many particles reach the surface of a planet per square meter, we integrate the above function from 1 Gev to $10^{12}$ Gev and in $2\pi$ sterradianes. We have that arrive to Earth

$$N = 3.48 \times 10^4 \text{ particles } m^{-2} s^{-1} \qquad (2)$$

It is expected that the cosmic rays flux on Mars is slightly greater than in Earth. But the precise proportion it is unknown. Here I suppose than the cosmic ray flux on Mars is similar to that on Earth. Mars has a radius of 3389.9 km, i.e., a surface $1.444 \times 10^{14}$ $m^2$. Multiplying the cosmic rays flow by the area of Mars we have that the amount of energetic particles that reach Mars per second are:

$$N_M = 5.03 \times 10^{18} \text{ } s^{-1} \qquad (3)$$

If this flow was constant throughout the history of the solar system, then in 4.6 Ga the amount of cosmic ray particles that have reached Mars are

$$N_T = 7.31 \times 10^{35} \qquad (4)$$

*2.2 Probability that a particle of cosmic rays collide with atomic nuclei of the atmosphere*
Suppose that a proton of cosmic rays, enters a prism of atmosphere with a base area A and height λ that is the mean free path of the proton. Inside this prism, there are N nuclei, which can impact the proton and produce fission. The probability of impact with a nucleus is

$$p = \frac{a}{A} \quad (5)$$

Where a is the sum of all cross sections of the N nuclei.
We have

$$a = \pi r^2 N \quad (6)$$

Where r is the radius of the nuclei of the material within the prism.
Also

$$N = A \lambda n \quad (7)$$

Where n is the number density of nuclei in the prism.
Substituting (7) in (6) and this in (5) we have

$$p = \pi r^2 \lambda n \quad (8)$$

On the other hand the mean free path is

$$\lambda = \frac{1}{\sigma n} \quad (9)$$

Substituting into equation (8) the value of λ given in equation (9) then we have the probability that the proton impact one of the nuclei is

$$p = \frac{\pi r^2}{\sigma} \quad (10)$$

For high energies the cross section is almost equal to $\pi r^2$ so the probability is almost 1, i.e., all protons of the primary cosmic rays, will impact against nuclei of the atmosphere, if the atmosphere is so dense as to have a thickness equal to, or greater than, a mean free path. The mean free path in an ideal gas at standard temperature and pressure is 11.85 km.

*2.3 Mass of lithium formed by cosmic rays on Mars*
To produce fission of atoms of the atmosphere, is needed the absorption of thermal neutrons with energy of MeVs. When a particle of high energy cosmic rays (> 1 GeV) hits atomic nucleus fission does not occur, but it breaks apart all nucleons (Pomerantz, 1971, p 82-84). The protons released in this destruction still have enough energy to break other nuclei. After two or three impacts of this nature, the original energy has been distributed among several particles and the neutrons from the last nuclear destruction have energies of the order of MeV and can produce fission reactions that generate lithium.

Lithium generated by cosmic rays

In an atmosphere composed mainly of $CO_2$, the impact of high-energy proton from space, produces 7 energetic protons if the impact is with a carbon (6 of the nucleus and the original proton) and 9 if the impact is with oxygen. Since there are two times more oxygen than carbon then, on average, 8.33 energetic protons are produced by each destructive interaction.

Neutrons produced in the case of carbon are 6, and in the case of oxygen 8, therefore on average 7.33 neutrons are produced by each destructive interaction.

The protons produced on the first interaction again interact with other nuclei and in the second interaction the number of thermal neutrons to be produced will be 8.33 x 7.33 = 61.06. I.e., for each energetic proton arriving from space, 61.06 slow neutrons are produced that can produce fission of the nuclei of carbon and oxygen and produces lithium.

In an atmosphere composed almost by pure $CO_2$, as those of Mars and Venus, the reactions that are possible between slow neutrons and atoms of carbon and oxygen are 15 in total. With carbon nuclei are 7, which only one produces lithium atoms, but it is the more probable:

$$^1_0n + ^6_6C \rightarrow ^7_6C* \rightarrow ^4_3Li + ^3_3Li \qquad (11)$$

The reactions that take place between slow neutrons and the nuclei of oxygen are 11, of these 4 produce lithium atoms and the 2 of maximum probability are:

$$^1_0n + ^8_8O \rightarrow ^9_8O* \rightarrow ^6_5B + ^3_3Li \qquad (12)$$

$$^1_0n + ^8_8O \rightarrow ^9_8O* \rightarrow ^5_5B + ^4_3Li \qquad (13)$$

The asterisk indicate that the nucleus considered is unstable because contains excess energy. The upper index indicates the number of neutrons and lower index indicates the number of protons.

As the carbon and the oxygen are in a proportion of 1:2, then from the 61.06 neutrons generated by a energetic proton, 20.35 interact with carbon, and 40.71 interact with oxygen.

If the reactions (11), (12), and (13) are much more probable than the other 12, then, the amount of mass of lithium, generated by cosmic rays is obtained as follows: Since the reactions with carbon produce two lithium atoms, then it is produced 40.70 nuclei of lithium. And since with the oxygen only one nucleus of lithium is produced, then 40.7 lithium nuclei are produced. In total, for each energetic proton that arrives from space are produced 81.4 lithium atoms.

If we multiply this number by the number of cosmic ray particles have reached Mars in its history (eq 4), then we have the number of lithium atoms produced by cosmic rays that we can expect to find on Mars

$$N_{Li} = 81.4 \times 7.31 \times 10^{35} = 5.95 \times 10^{37} \qquad (14)$$

Lithium generated by cosmic rays

During the process are produced atoms of lithium 7 and 6 in equal proportion. Therefore, we can assume an average atomic weight of 6.5. Taking this value, the mass of the metal produced in the history of Mars is:

$$M_{Li} = 6.5 \text{ amu} \times 1.66 \times 10^{-27} \text{ kg/amu} \times 5.95 \times 10^{37} = 6.42 \times 10^{11} \text{ kg} \qquad (15)$$

This is 642 million metric tons.
On the other side, if the 15 possible reactions between the neutrons and the carbon and oxygen atoms have equal probability of occur then the mass of lithium would be estimated as follows:
As the carbon and the oxygen are in a proportion of 1:2, then from the 61.06 neutrons generated by a energetic proton, 20.35 interact with carbon, and 40.71 interact with oxygen.
Of the seven possible reactions between neutrons and carbon, only one forms lithium and produces 2 atoms. If the seven reactions are equally probable, then 2/7 of the 20.35 neutrons form lithium, i.e., are produced 20.35 x 2/7 = 5.81 lithium atoms.
With oxygen, there are 11 possible reactions, which together generate 4 lithium atoms. That is, the number of lithium atoms generated, if all reactions have the same probability to occur, will be 40.7 x 4/11 = 14.8.
In other words, for each proton of cosmic rays hitting the planet, are formed 20.61 lithium atoms, if all 18 possible reactions with carbon and oxygen are equally likely.
Multiplying by the number of protons received by Mars (equation 4) we have

$$N_{Li} = 20.61 \times 7.31 \times 10^{35} = 1.5 \times 10^{37} \qquad (16)$$

And the total mass is

$$M_{Li} = 6.5 \text{ amu} \times 1.66 \times 10^{-27} \text{ kg/amu} \times 1.5 \times 10^{37} = 1.62 \times 10^{11} \text{ kg} \qquad (17)$$

This is 162 million metric tons.
The amount of lithium mass generated on Mars by cosmic rays depends on the probability of each of the reactions between the neutrons and the nuclei of the atmosphere but is between 162 and 642 million metric tons.
The estimated reserve of lithium in the Earth is 30 million metric tons (Lagos-Miranda, 2009). I.e. Mars has, only by interaction with cosmic rays, between 5.4 and 21.4 times that Earth.
This amount of mass is obtained assuming that in all the geological history of Mars the atmosphere had a pressure high enough so that cosmic rays have had at least two energetic collisions with atmospheric atoms (primary and secondary protons), to generate thermal neutrons that allow the fission of atoms of carbon and oxygen. For this the mean free path of the cosmic rays protons must be at most half of the scale height of the atmosphere. To a pressure of one atmosphere the mean free path is 11.85 km, of the same order of the height scale on Mars, which is currently 11 km (Tholen et al, 2000). Therefore, to produce lithium, the atmospheric pressure should be at least 2 bars. And so the true amount of lithium produced by cosmic rays on Mars will be $M_{Li}\,t/T$ where t is the time that Mars had a thick atmosphere (> 2 bars) and T are the 4.6 Ga of its history. In this way lithium may be used as an indicator of the amount of time in which Mars had an atmosphere of more than two bars.

*3 Discussion*

Since cosmic rays come isotropically into the atmosphere of Mars then we can assume that lithium formed by cosmic rays will be distributed homogeneously across the surface. If this were the case would have 4.427 kg/km² t/T. Assuming that this amount is distributed in a stratigraphic column of 5 km depth then we have a density of 885.4 kg/km³ t/T, or 8.854x10-7 kg/m³ t/T. This would be the density of lithium that the rover "Curiosity" will find on the surface on Mars if there are no mechanisms that facilitate its concentration.

However lithium compounds are soluble in water then tends to concentrate on it, which as it evaporates, can form lithium-rich evaporites. On Earth about 60% of lithium production is obtained from brines (Lagos-Miranda, 2009). If in the past Mars has had lakes, rivers and oceans could be formed lithium-rich deposits. This would be a reason for mining the mineral on the planet. On the other side from the astrobiological point of view to find lithium-rich evaporites on Mars would be an indication that there was liquid water, a condition for the emergence of life.

If in the past there was liquid water on Mars, lithium formed by cosmic rays in the atmosphere react with water to form lithium hydroxide, as follows (Lenntech, 2012)

$$2Li + 2H_2O \rightarrow LiOH + H_2 \qquad (16)$$

Furthermore lithium hydroxide reacts with $CO_2$ to form lithium carbonate, as follows (EcuRed, 2012)

$$2LiOH + CO_2 \rightarrow Li_2CO_3 + H_2O \qquad (17)$$

From these two reactions are expected that Mars has deposits of lithium carbonate, if there was liquid water in the past.

This study did not consider the planet's original lithium, which acquired during accretion, since it is very likely to be trapped in the planetary mantle due to the lack of plate tectonics, and just a little came to the surface due to volcanism. For this reason, most of the lithium found in the crust of Mars must be generated by cosmic rays.

*4 Conclusions*

The conclusions of this work are:

Lithium, produced by cosmic rays, can be found on Mars, either homogeneously distributed throughout the surface, or in deposits.

Find lithium, although not in deposits, tell us how long Mars had an atmosphere with more than 2 bars, which is an important factor in astrobiology potential of the planet.

On the other hand the importance of finding lithium deposits on Mars is twofold. On the one hand we would say that it exists or existed, liquid water in the planet's surface.

Lithium generated by cosmic rays

This in turn is an important factor for the possible emergence and evolution of life on Mars. Moreover, another consequence of importance of finding lithium deposits on Mars would be economic, because it could be exploited these deposits and bring the metal to Earth. This is feasible from a technological standpoint but probably would be too expensive to be profitable.

*References*

Clayton, D. (2003) Isotopes in the Cosmos. Hydrogen to Gallium. Cambridge University Press.

Dorman, L. (2006). Cosmic ray interactions, propagation, and acceleration in space plasmas. Ed. Springer.

EcuRed (http://www.ecured.cu/index.php/Hidróxido_de_litio), (Consulted June 15, 2012)

Lagos-Miranda, C. (2009). Antecedentes para una política publica en minerales estratégicos: litio. Comisión Chilena del Cobre, Dirección de Estudios y Políticas Publicas. http://www.cochilco.cl/productos/pdf/2009/informe_minerales_estrategicos_litio.pdf (Consulted June 15, 2012).

Lenntech, B.V. (http://www.lenntech.es/litio-y-agua.htm), (Consulted June 15, 2012)

Pomerantz, M.A. (1971). Cosmic Rays. Van Nostrand Reinhold Company.

Tholen, D.J., V.G. Tejfel and A.N. Cox. 2000. Chapter 12 Planets and Satellites. In Allen's Astrophysical Quantities. 4th Edition Editor A.N. Cox. AIP Press and Springer.